# INTERFACE

**Author for correspondence:**
Valerio Capraro
e-mail: caprarovalerio@gmail.com

# Language-based game theory in the age of artificial intelligence

Valerio Capraro[1], Roberto Di Paolo[2], Matjaž Perc[3,4,5,6] and Veronica Pizziol[7]

[1]Department of Psychology, University of Milan Bicocca, Milano, Italy
[2]Department of Economics and Management, University of Parma, Parma, Italy
[3]Faculty of Natural Sciences and Mathematics, University of Maribor, Maribor, Slovenia
[4]Community Healthcare Center Dr. Adolf Drolc Maribor, Maribor, Slovenia
[5]Complexity Science Hub Vienna, Vienna, Austria
[6]Department of Physics, Kyung Hee University, Seoul, Republic of Korea
[7]Department of Economics, University of Bologna, Bologna, Italy

VC, 0000-0002-0579-0166; RDP, 0000-0002-6081-6656; MP, 0000-0002-3087-541X; VP, 0000-0002-9032-8506

Understanding human behaviour in decision problems and strategic interactions has wide-ranging applications in economics, psychology and artificial intelligence. Game theory offers a robust foundation for this understanding, based on the idea that individuals aim to maximize a utility function. However, the exact factors influencing strategy choices remain elusive. While traditional models try to explain human behaviour as a function of the outcomes of available actions, recent experimental research reveals that linguistic content significantly impacts decision-making, thus prompting a paradigm shift from outcome-based to language-based utility functions. This shift is more urgent than ever, given the advancement of generative AI, which has the potential to support humans in making critical decisions through language-based interactions. We propose sentiment analysis as a fundamental tool for this shift and take an initial step by analysing 61 experimental instructions from the dictator game, an economic game capturing the balance between self-interest and the interest of others, which is at the core of many social interactions. Our meta-analysis shows that sentiment analysis can explain human behaviour beyond economic outcomes. We discuss future research directions. We hope this work sets the stage for a novel game-theoretical approach that emphasizes the importance of language in human decisions.

## 1. Introduction

Understanding human behaviour in decision problems and strategic interactions has long been at the heart of social science research due to its wide-ranging applications in various fields such as economics, psychology and artificial intelligence. In the past century, game theory has provided a compelling framework for this understanding, underpinned by the notion that individuals seek to maximize a utility function [1]. Yet, the components of this utility function remain unknown. What factors do people consider in their strategy choices?

One of the central tools for this exploration has been economic games [2]. These games are particularly insightful for understanding human actions in situations where personal interests are at play. In particular, the study of one-shot and anonymous games, where participants interact once without the possibility of future interactions and without possessing any information about the others, has garnered particular attention, as they offer a clean benchmark to study human behaviour, free from the confounding factors of direct and indirect reciprocity.

It has long been known that even in these one-shot and anonymous games, individuals often do not act purely to maximize their economic benefits. Take,

**THE ROYAL SOCIETY** PUBLISHING

for instance, the dictator game. In this game, a participant has to decide how to divide a sum of money between themselves and another player. The other player has no active role and only receives the amount that the first player decides to give. This is one of the most studied games in behavioural economics, due to its ability to capture how people balance self-interest (keeping all the money) and the interest of others (giving away some of the money). Despite the absence of extrinsic incentives, a significant number of participants choose to share some amount [3–5]. This brings us to one of the most fundamental questions in behavioural game theory: if not monetary gain, what exactly are people optimizing for?

## 2. A paradigm crisis

Over the past 30 years, the concept of 'social preferences' has gained traction. Here, an individual's utility is a function not just of their own monetary outcomes but also those of others with whom they are interacting. The formalization of this idea has taken various shapes (e.g. [6–11]; see [12,13] for reviews). Ledyard, for example, postulates that people combine a preference for maximizing their own monetary payoff with a preference for maximizing the payoff for all others involved [6]. By contrast, Fehr and Schmidt's utility revolves around the idea that people aim to maximize their own monetary rewards but also strive to minimize the disparity between their gains and those of other players [8]. Charness and Rabin take a different route, assuming that people aim to optimize their own benefits while simultaneously maximizing the overall welfare of the group [10]. Therefore, these models, though distinct, share a central 'consequentialist assumption': the utility derived from a decision depends solely on the economic outcomes of that choice.

However, this consequentialist assumption has recently been subjected to serious criticism. One major point of contention arises from experimental research with human participants, emphasizing the profound influence of linguistic content on decision-making. Simply put, the way actions are described can significantly alter people's choices, challenging the purely consequentialist assumption of social preferences. In a pioneering study, Liberman and colleagues observed that mere linguistic labels could affect individuals' behaviour in the Prisoner's Dilemma. Specifically, when a Prisoner's Dilemma game was labelled as a 'community game', participants tended to cooperate more often compared to when the very same game was labelled as a 'Wall Street game' [14]. This linguistic impact on decision-making was further reinforced by Eriksson and colleagues. Their research revealed that responders in the ultimatum game were more prone to rejecting low offers when the rejection action was phrased as 'reject the proposer's offer', as opposed to 'decrease the proposer's payoff' [15]. Similarly, a study by Capraro and Rand identified that linguistic framing influenced decisions involving equity versus efficiency trade-offs. When the two options were respectively termed 'more fair' versus 'less fair', participants leaned towards the equitable choice. However, when framed as 'less generous' versus 'more generous', participants were inclined towards the efficient option [16]. Furthermore, Capraro and Vanzo demonstrated the power of language in the dictator game. They conducted six dictator game variants differing only in the label used to describe the available actions and found that people's level of altruistic behaviour significantly depended on the label. For example, individuals were less inclined towards altruism when the altruistic action was labelled as 'boost' rather than 'donate' [17]. In the last 5 years, the effects of language on decisions have been replicated many times in different contexts [18–24]. Some research has also highlighted a dark side of the linguistic framing effect. For example, Capraro and colleagues have shown that when dictator game receivers are given the power to choose the experimental instructions to present to dictators, a significant proportion of receivers choose instructions that are more likely to provide them a higher payoff [25]. Furthermore, Ścigała and colleagues have found that more moral people—defined as those higher in the personality trait of honesty-humility—can be manipulated and turned into accepting a bribe by simply calling the bribe a 'cooperation act' [26].

Cumulatively, these studies challenge the consequentialist assumption of social preferences and demonstrate that utility functions cannot be merely based on the economic outcomes of the available actions. Utility functions must take into account the language used to describe the actions. For this reason, it has been argued that 'behavioural economics is in the midst of a paradigm shift from outcome-based to language-based preferences' [13].

## 3. The importance of language-based preferences in human–machine interactions

This paradigm shift is more urgent than ever due to the rise of generative artificial intelligence (AI). The evolution of generative AI is reshaping our digital landscapes in ways previously thought impossible. One of the most transformative aspects of this revolution is the increasing capability of AI systems to generate human-like, coherent and contextually relevant textual content. OpenAI's generative pre-trained transformer (GPT) series, for instance, exemplifies this shift, showcasing the ability to produce text that not only reads naturally but also responds contextually to diverse prompts [27].

This rise of text-generating AI has profound implications for decision-making support. As AI becomes more sophisticated, chatbots and virtual assistants powered by these algorithms can guide users in making critical decisions [28–30]. Whether it is a financial choice, a health concern or a complex business strategy, AI chatbots can present users with detailed analyses, recommendations and even potential consequences, all communicated in natural language.

However, these potential benefits come with critical challenges. Previous literature has identified several potential issues, such as accountability, especially when AI-guided decisions lead to adverse outcomes, and the risk of overreliance, which could erode human judgement and decision-making abilities [31–33]. In this article, we shift focus to an often-overlooked aspect: the linguistic description of the decision context. As reviewed above, even in the simplified case of one-shot and anonymous economic games, human decision-making is not just a by-product of the economic consequences of the available actions; it is heavily influenced by the way information is presented. Considering AI's capacity to generate text, it is therefore vital to understand and account for these linguistic frames. If AI chatbots are to aid in important decisions, they need to be designed with an awareness of the effects of language framing. Misrepresentation or linguistic biases, even if unintentional, could lead individuals down undesired paths, with potential downstream negative effects at the social level. For example, linguistic biases may exacerbate discrimination against marginalized groups [34–36]. Moreover, there is also an ethical imperative to ensure that AI's capability to produce text is not misused, manipulating users' decisions to serve ulterior motives [37–39].

Closing the circle and going back to a game-theoretical point of view, this set of concerns brings us to a novel question: how can the linguistic content of a piece of text be quantified in a way that can be incorporated into the utility function?

# 4. Using sentiment analysis to define the utility function over language

A straightforward idea is to use sentiment analysis [40]. Sentiment analysis, at its core, is a set of tools developed by computational linguists to evaluate the emotional tenor of a given piece of text. These tools, driven by complex algorithms and based on large linguistic databases, scan textual information to identify and quantify the emotional content embedded within it. Earlier sentiment analysis tools were primarily limited to determining whether a text was positive or negative without taking into account different human emotions. For example, SentiWordNet associates with each 'synset' (a set of synonyms) three continuous numeric scores: positivity, negativity and neutrality, which together sum up to 1. To evaluate the sentiment of a text using SentiWordNet, an algorithm breaks the text into constituent terms or phrases. Each term's corresponding synset scores are then calculated through a suitable automatic annotation process. By aggregating these scores across the entire text, an overall sentiment value can be derived. The aggregation involves weighted sums, where more contextually important words have a greater influence on the final sentiment [41]. More recently, newer sentiment analysis tools have begun to emerge. These tools aim to go beyond the basic binary of positivity and negativity. For example, Mohammad and Turney developed a tool that can identify and measure a spectrum of emotions, such as joy, sadness, anger, fear, surprise, and disgust. By mapping words to specific emotions and emotional intensities, this tool offers a more faceted understanding of sentiment in textual data, capturing the multidimensional nature of human emotions more effectively than earlier models [42].

Yet, while understanding and measuring emotions in the text is undeniably valuable, it is essential to consider that there is more to text sentiment than just emotions. Some linguistic content might emphasize specific behavioural norms or cultural standards, which might be relatively detached from the emotions they convey. Therefore, sentiment analysis tools need to be extended or adapted to measure the normative content of text, detached from emotions. Some work has already been done on this. Moral foundations theory offers one framework for such analysis, emphasizing five (later extended to six) core moral values: care, fairness, loyalty, authority, sanctity and liberty [43]. Computational models have been developed to identify these moral values in textual data, effectively giving a 'normative score' based on these foundations [44]. The same happens for the more recent morality-as-cooperation theory [45].

Leaving aside this intricate web of emotional and normative dimensions, if we were to simplify our approach, the general idea would be to use sentiment analysis tools to quantify the linguistic descriptions of possible actions in a decision problem. These quantified descriptions, represented as numerical scores, would then be fed into a utility function. Of course, implementing such a procedure is not straightforward. The intertwining of sentiment analysis with utility functions would invariably be complex. However, there might be exceptions. For economic games with clear-cut decisions, such as the dictator game, the integration might be more straightforward due to the inherent simplicity of the decision structure in such scenarios.

# 5. Predictions in the dictator game

Meta-analytic results of the available experimental literature have revealed that most players in the dictator game choose one of three actions: keeping all the money, sharing it equally, or giving it all away [5]. In an attempt to simplify our proposed approach, we can use these results to assume that when participants in the dictator game make their decision, they consider the utility of only these three actions, disregarding other possible actions.

In our proposed model, the utility that a player derives from choosing a particular strategy is determined by two factors: the monetary payoff and the sentiment associated with the description of that strategy.

Considering material payoffs, the strategy yielding the highest monetary return is to keep all the money. This is followed by sharing half (which provides a moderate return) and then giving away all the money (which results in no material gain). However, when we introduce the concept of sentiment into the equation, the analysis becomes more complex, as the sentiment associated with each action can significantly influence the player's decision. There are three cases:

— If the sentiment attached to acting selfishly (keeping all the money) is the highest, a player will choose this option, as it maximizes both material payoff and sentiment.
— If the sentiment for the inequity-averse action (sharing the money equally) is the highest, the player faces a dilemma between the selfish action, which offers the highest material payoff, and the egalitarian action, which aligns with the most favourable sentiment. In this case, acting altruistically (giving all the money) is disregarded, as it is dominated by the inequity-averse action, both in terms of material payoff and sentiment.
— If the sentiment for the altruistic action is the highest, the player experiences a tension among all three available actions. The strength of the sentiment towards the altruistic action directly influences the likelihood of its selection. Similarly, a strong sentiment towards the inequity-averse action increases the propensity to choose this over the other options. If we label actions as 'prosocial' when they are either inequity-averse or altruistic, we can deduce that the higher the average sentiment between these two actions, the more likely a player is to make a prosocial choice.

Therefore, even without pinpointing the precise utility function, we can derive a testable hypothesis. Let us define

$$\Delta S = \begin{cases} S_{\text{half}} - S_{\text{zero}} & \text{if } S_{\text{all}} \leq S_{\text{half}} \\ \frac{S_{\text{all}} + S_{\text{half}}}{2} - S_{\text{zero}} & \text{otherwise,} \end{cases}$$

where $S_{\text{half}}$ is the sentiment score associated with the action of 'giving half of the endowment', $S_{\text{all}}$ is the sentiment score associated with 'giving all the endowment' and $S_{\text{zero}}$ is the sentiment score associated with 'keeping all the endowment'. Then, we obtain the following:

*Hypothesis: $\Delta S$ is positively correlated with the rate of prosocial behaviour in the dictator game.*

If this hypothesis holds, it could pave the way for more intricate game-theoretical and decision-making models where sentiment-driven language-based utility functions play a pivotal role.

## 6. A meta-analysis of dictator game experiments

To test this hypothesis, we issued a public call on the forums of the Economic Science Association (ESA) and the Society for Judgment and Decision Making (SJDM), asking behavioural scientists to provide instructions of dictator game experiments with human participants they had conducted. We supplemented this call with manual searches of the relevant literature with the aim of collecting as many experimental instructions as possible.

Since different experimental studies may differ in many dimensions other than language (e.g. nationality of the sample, gender balance, age; all these variables have been shown to affect behaviour in the dictator game [5]), our aim is to calculate the values $S_{\text{zero}}$, $S_{\text{half}}$ and $S_{\text{all}}$ at the study level and then use these values to predict the rate of altruistic behaviour as a function of $\Delta S$ in each single study using a linear regression. Then, we will use the coefficients and standard errors of the study-level regressions to conduct a meta-analysis of all the studies. Meta-analysis is a statistical technique that synthesizes data from multiple studies to identify patterns, trends and overall effects, while also measuring heterogeneity across studies (see [46–48] for examples of meta-analyses on similar games). To estimate coefficient and standard error at the study level using a linear regression, we need studies with at least three experimental conditions because a linear regression with two data points returns no standard error since there is only one line that passes through two distinct points. We collected 12 research articles with a total of 61 experimental conditions that satisfy this requirement [17,24,49–58].

Following Rathje and colleagues [59], we employed the generative AI chatbot GPT-4 to conduct sentiment analysis. In their study, Rathje and colleagues demonstrated that GPT-4 accurately detects sentiments close to fine-tuned machine learning models. We adapted their procedure to our setting, and we prompted GPT-4 to evaluate the sentiment associated with the three prominent actions of the dictator game, namely, 'keeping all the endowment', 'keeping half of the endowment' and 'giving all the endowment'. To do so, each instruction was inputted into the chatbot with the following prompt: 'Now imagine that there is a population of 1000 people living in [country]. What do you think the average response to the following questions would be? (Please return an exact number with two decimal digits). How negative or positive is the action of [action referring to 'keeping all/keeping half/giving all the endowment'] on a 1–7 scale, with 1 being 'very negative' and 7 being 'very positive'?'. In this prompt, [action referring to . . .] was substituted with the exact words used by the authors in the specific experimental instruction, and [country] was replaced with the country where the experiment was conducted. The box below reports an example of the prompt.

We recorded numerical responses for each experimental instruction. The output of this methodology provided a sequence of AI-generated sentiment scores associated with each prominent action, which we then used to calculate the value $\Delta S$. To prevent any learning and maintain the integrity of the analysis, we deleted the chat with GPT-4 after collecting the score for each instruction, ensuring that subsequent estimations were not influenced by previous conversations.

Table 1 reports the average sentiment score associated with each prominent action in dictator games.

On average, the sentiment score associated with the pro-self action is lower than the sentiment score associated with the prosocial actions. Moreover, the sentiment score associated with the altruistic action (i.e. 'giving all') is similar to that of the inequity-averse action (i.e. 'giving half'). However, looking at the disaggregated data, in some cases, GPT-4 returned a higher sentiment score for the altruistic action, while in other cases, it returned a higher sentiment score for the inequity-averse action, underscoring the importance of keeping the two cases separated, as done in the definition of $\Delta S$. Not surprisingly, instead, the self-regarding action got consistently lower rates. See table 2 for the disaggregated data.

Then, we turn to the meta-analysis. Figure 1 reports the forest plot of the random-effects meta-analysis. Forest plots provide the standard way to report meta-analytic results, as they represent a complete overview of the effect sizes and confidence intervals of individual studies while allowing for the assessment of an overall summary estimate while not losing track of heterogeneity. Random-effects meta-analysis is generally preferred over fixed-effects meta-analysis in situations, like ours, where there is heterogeneity among the studies being analysed. This heterogeneity can be due to differences in study populations, methodologies, interventions or other factors that might influence the outcomes. We mention, for completeness, that the results are robust and become actually stronger when using fixed-effects meta-analysis. Therefore, the random-effects estimation represents a conservative estimation. Coming to the meta-analytic results, in line with the hypothesis, we find a significantly positive effect (overall effect size = 0.08; 95% CI = [0.04, 0.13]; $z = 3.41$; $p < 0.01$), such that higher $\Delta S$ is associated with more prosocial behaviour. We have conducted several robustness checks. When asking GPT-4 to predict the average response of 1000 people in the USA (selected because most of the training dataset originated there), the outcomes are consistent (overall effect size = 0.09; 95% CI = [0.01, 0.16]; $z = 2.22$; $p = 0.03$). Similarly, the results are stable when not specifying the sample size and requesting GPT-4 to estimate the average response of a population living in [country] (Overall effect

**Table 1.** Descriptive statistics of the sentiment score associated with the prominent actions in dictator games.

| | $S_{zero}$ | $S_{half}$ | $S_{all}$ |
|---|---|---|---|
| average | 2.600 | 5.233 | 5.369 |
| s.d. | 0.627 | 0.929 | 1.010 |

**Table 2.** Descriptive statistics of the sentiment score associated with the prominent action in dictator games by experimental instructions.

| experimental instruction | $S_{zero}$ | $S_{half}$ | $S_{all}$ | experimental instruction | $S_{zero}$ | $S_{half}$ | $S_{all}$ |
|---|---|---|---|---|---|---|---|
| Antinyan et al. [50] | | | | Kettner & Ceccato [51] | | | |
| control | 3.20 | 5.50 | 4.75 | give female | 4.50 | 6.00 | 5.50 |
| loss manipulation 1 | 3.50 | 5.50 | 4.50 | give male | 4.50 | 6.00 | 5.25 |
| loss manipulation 2 | 2.50 | 5.00 | 4.00 | take female | 2.00 | 3.50 | 5.50 |
| Brañas-Garza [52] | | | | take male | 2.50 | 3.50 | 5.50 |
| baseline | 2.50 | 5.50 | 4.50 | Kettner & Waichman [53] | | | |
| helping others | 2.00 | 6.00 | 5.00 | give hypothetical | 4.50 | 5.50 | 5.50 |
| reciprocity | 2.50 | 6.50 | 4.50 | give incentivized | 4.00 | 5.50 | 5.00 |
| Bruttel & Stolley [54] | | | | take hypothetical | 2.00 | 2.50 | 6.00 |
| control | 3.00 | 5.50 | 4.50 | take incentivized | 2.00 | 2.50 | 5.50 |
| decision power condition | 2.50 | 5.50 | 4.50 | Kuang & Bicchieri [24] | | | |
| responsibility condition | 2.50 | 5.50 | 4.50 | control | 2.75 | 5.50 | 6.50 |
| Caparro & Vanzo [17] | | | | ind. injunction, appropriate | 3.25 | 5.50 | 6.75 |
| boost condition | 3.25 | — | 5.75 | ind. injunction, approved | 2.50 | 5.50 | 6.50 |
| demand condition | 2.25 | — | 5.75 | ind. injunction, desirable | 2.50 | 6.00 | 6.50 |
| donate condition | 3.25 | — | 5.75 | ind. injunction, okay | 3.25 | 5.75 | 6.50 |
| give condition | 2.75 | — | 5.75 | ind. injunction, permissible | 3.50 | 5.50 | 6.50 |
| steal condition | 1.50 | — | 6.50 | ind. injunction, should | 2.30 | 5.20 | 6.75 |
| take condition | 2.25 | — | 5.75 | ind. injunction, the right thing | 2.75 | 5.50 | 6.50 |
| Dreber et al. [55] | | | | Ockenfels & Werner [49] | | | |
| Exp 1 - giving informed | 2.15 | 5.40 | 6.20 | info condition 1 | 2.50 | 5.50 | 4.50 |
| Exp 1 - giving uninformed | 2.50 | 5.75 | 6.25 | info condition 2 | 2.50 | 5.50 | 4.50 |
| Exp 1 - taking informed | 1.50 | 2.50 | 6.00 | noInfo condition 1 | 2.50 | 5.50 | 4.00 |
| Exp 1 - taking uninformed | 1.75 | 4.25 | 6.50 | noInfo condition 2 | 2.50 | 5.50 | 4.50 |
| Exp 2 - giving give | 2.50 | 5.25 | 6.75 | Schurter & Wilson [56] | | | |
| Exp 2 - giving transfer | 2.50 | 5.25 | 6.25 | die roll condition | 2.50 | 6.00 | 4.50 |
| Exp 2 - keeping keep | 2.45 | 4.50 | 6.25 | quiz condition | 2.50 | 6.00 | 2.50 |
| Exp 2 - keeping transfer | 2.50 | 5.00 | 6.00 | seniority condition | 2.35 | 6.45 | 4.20 |
| Exp 3 - giving Informed | 2.15 | 5.45 | 6.50 | unannounced condition | 2.45 | 6.10 | 2.75 |
| Exp 3 - giving uninformed | 2.50 | 5.00 | 6.50 | Walkowitz [57] | | | |
| Exp 3 - taking informed | 2.50 | 4.00 | 6.50 | DeRo25 | 2.50 | 5.50 | 4.50 |
| Exp 3 - taking uninformed | 2.15 | 3.65 | 6.20 | Dec50 | 2.25 | 5.75 | 4.75 |
| Herne et al. [58] | | | | N-N | 2.00 | 5.00 | 4.00 |
| baseline | 2.50 | 5.50 | 5.00 | N-N-2 | 2.50 | 4.50 | 5.50 |
| certainty empathy | 2.00 | 5.80 | 4.80 | Pay50 | 2.50 | 5.50 | 4.50 |
| uncertainty empathy | 2.50 | 6.00 | 5.00 | Rol50 | 2.50 | 5.50 | 4.50 |
| uncertainty no empathy | 2.20 | 5.80 | 4.40 | | | | |

size = 0.08; 95% CI = [0.00, 0.15]; $z = 2.07$; $p = 0.04$). Finally, the results maintain their robustness when prompting GPT-4 with all the conditions from a given study, without restarting the chat after each condition (Overall effect size = 0.08; 95% CI = [0.03, 0.12]; $z = 3.60$; $p < 0.01$).

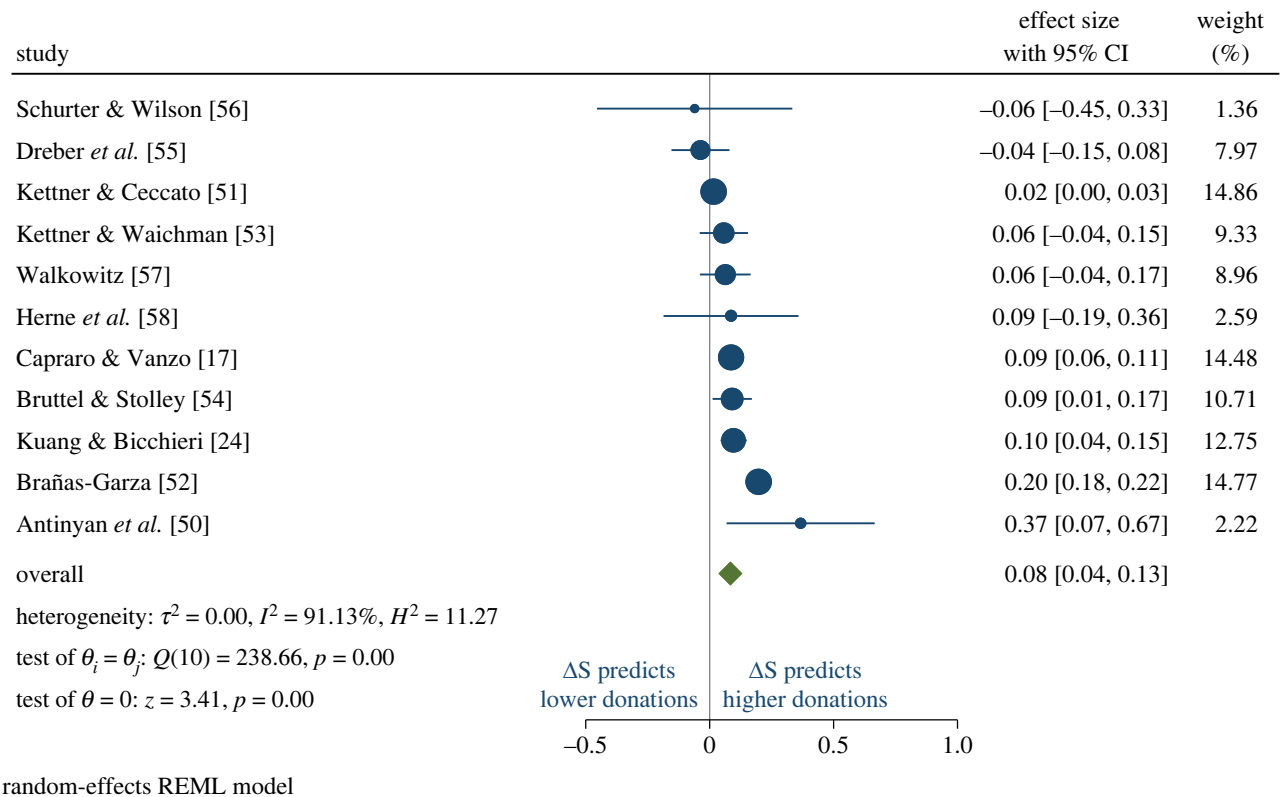

**Figure 1.** Forest plot of the meta-analysis of the sentiment associated with altruistic behaviour across the studies. One study [49] gets dropped from the meta-analysis, because GPT-4 estimates the same sentiments in all four conditions, therefore the study-level regression cannot estimate the standard error. It is important to note that GPT-4 sentiment scores are in line with actual behaviour also in this study, as the mean giving did not significantly vary across conditions. Moreover, the same study does not get dropped in the robustness checks, all of which show a similar pattern of results.

# 7. Discussion

We are entering an era marked by the rapid advancement of artificial intelligence. The implications of this evolution are bound to be transformative, altering the very fabric of how we live, work and think. The promise of this new age is not just in the computational power of these machines but in the profound ways they are expected to blend into the human experience [60,61]. Among other things, the synergy between humans and machines holds the potential to amplify our decision-making processes. Machines, free from the cognitive limitations that humans face, may guide us to make more informed, optimal decisions, especially in critical situations where human judgement may vacillate due to stress [62], cognitive overload [63], or bias [64].

But the benefits of this symbiotic relationship come with their own set of challenges [31]. In this article, we paid attention to one particular challenge. Since humans and machines will primarily communicate through language, it becomes crucial to ensure that machines understand and process language in a way that aligns with human intent. From a game-theoretical viewpoint, this necessitates a paradigm shift from utility functions based solely on outcomes to utility functions that consider the influence of language on human preferences. Recognizing this, we advocate for the incorporation of sentiment analysis as a tool to develop language-based utility functions. Sentiment analysis could provide a quantitative measure of the language that describes possible actions in decision-making scenarios.

We have embarked on this path by leveraging the sentiment analysis capabilities of GPT-4 to shed light on human choices within the context of the dictator game. This game is emblematic of the complex interplay between self-interest and the interest of others, capturing the essence of a multitude of social interactions where one must balance personal gain against the welfare of others [5].

Our findings serve as a starting point, indicating the potential of sentiment analysis in explaining human decisions. Yet, this is merely the first step of a broader exploration. Future work should extend the application of sentiment analysis to other spheres of human interactions, including cooperation [65], honesty [66], altruistic punishment [67], trust and trustworthiness [68]. Moreover, the binary paradigm of classifying sentiments into positive or negative categories must be overcome. It is imperative to account for a spectrum of emotions and moral values, each with distinct influences on behaviour, as experimental studies have shown that different emotions can sway decisions in various ways [69–71], just as diverse ethical considerations can steer actions along different paths [72–74]. Building on the methodology outlined in this article, an initial step could be to test GPT-4's ability to reliably assess various emotions and normative dimensions, and to determine how these assessments might explain behavioural patterns. Alternatively, refined content analysis tools could be used or developed for this purpose. For instance, several sentiment analysis tools have been developed to identify and quantify a large set of emotions [42,75], while tools for identifying and quantifying diverse moral values have started to emerge in recent years [76,77]. Furthermore, the exploration should not stop at the opaque algorithms of GPT-4. The predictive capacity of various sentiment analysis tools should be examined, dissecting the 'black box' to understand the mechanics of language's influence on decision-making. This will be fundamental for developing and testing a specific utility function that mathematically formalizes the complex interplay between language, emotions, norms and behaviour. Finally, the integration of sentiment analysis and language-based utility functions into game theory, as proposed in this study, opens up exciting possibilities for its combination with evolutionary game theory and mathematical modelling. Evolutionary game theory has significantly contributed to understanding the development of moral behaviours, such as cooperation [78–81], trust and trustworthiness [82], and honesty [83,84]. By incorporating sentiment analysis into the modelling of player strategies, we can capture a more nuanced understanding of human decision-making behaviour that extends beyond outcome-based preferences. For example, the use of sentiment analysis could help model the evolution of cooperation or honesty in a population, where the language used to describe actions could influence the perceived utility of those actions and thus the evolution of strategies. Mathematically, these language-based utility functions could be integrated into the replicator dynamics equations used in evolutionary game theory, adding a new dimension to these models.

In summary, our aspiration is that this article lays the groundwork for a novel approach in game theory, one that recognizes the importance of language in decision-making processes. The journey ahead is filled with important questions necessitating dedicated research. The answers we find, and the questions we ask, will shape the future not only of game-theoretical research but also of the very nature of the relationship between humans and machines.

**Ethics.** This work did not require ethical approval from a human subject or animal welfare committee.

**Data accessibility.** Data and analysis code are available from the OSF repository: https://osf.io/mx5w3/ [85].

**Declaration of AI use.** Yes, we have used AI-assisted technologies in creating this article.

**Authors' contributions.** V.C.: conceptualization, data curation, investigation, methodology, writing—original draft, writing—review and editing; R.D.P.: data curation, formal analysis, methodology, writing—review and editing; M.P.: conceptualization, supervision, writing—review and editing; V.P.: data curation, formal analysis, methodology, writing—review and editing.

All authors gave final approval for publication and agreed to be held accountable for the work performed therein.

**Conflict of interest declaration.** We declare we have no competing interests.

**Funding.** M.P. was supported by the Slovenian Research and Innovation Agency (Javna agencija za znanstvenoraziskovalno in inovacijsko dejavnost Republike Slovenije) (grant nos. P1-0403 and N1-0232).

**Acknowledgements.** We thank Redi Elmazi for assistance during materials collection and the participants of the BEE meeting at the IMT School for Advanced Studies Lucca and of the 11th BEEN meeting at University of Bologna for their comments. We are grateful to the behavioural scientists who responded to our call on the ESA and SJDM forums and provided their experimental instructions.